\documentclass[10pt, a4paper, twocolumn]{article}

\usepackage[
    citestyle=numeric-comp,
    backend=biber,
    bibencoding=inputenc,
    urldate=long
    ]{biblatex}
\usepackage{authblk}
\usepackage[centering,margin=1.5cm]{geometry}
\usepackage[T1]{fontenc} 

\usepackage{graphicx}
\usepackage{textcomp}
\usepackage{xcolor}
\usepackage[utf8]{inputenc}
\usepackage[english]{babel}
\usepackage[autostyle=true]{csquotes}
\usepackage{url}

\usepackage{footnote}
\usepackage{tabularx}
\usepackage{multirow}
\usepackage{booktabs}
\usepackage{comment}
\usepackage[capitalise, noabbrev]{cleveref}

\usepackage{amssymb}

\begin{document}

  \title{MockFog 2.0: Automated Execution of Fog Application Experiments in the Cloud \thanks{This work has been published in IEEE Transactions on Cloud Computing. \textsuperscript{\textcopyright}2021 IEEE.}}

\author{Jonathan Hasenburg, Martin Grambow, and David Bermbach}
\affil{Mobile Cloud Computing Research Group,\\ TU Berlin \& Einstein Center Digital Future \\\{jh,mg,db\}@mcc.tu-berlin.de}

\maketitle

\begin{abstract}
Fog computing is an emerging computing paradigm that uses processing and storage capabilities located at the edge, in the cloud, and possibly in between.
Testing and benchmarking fog applications, however, is hard since runtime infrastructure will typically be in use or may not exist, yet.
While approaches for the emulation of infrastructure testbeds do exist, their focus is typically the emulation of edge devices.
Other approaches also emulate infrastructure within the core network or the cloud, but they miss support for automated experiment orchestration.

In this paper, we propose to evaluate fog applications on an emulated infrastructure testbed created in the cloud which can be manipulated based on a pre-defined orchestration schedule.
Developers can freely design the infrastructure, configure performance characteristics, manage application components, and orchestrate their experiments.
We also present our proof-of-concept implementation MockFog 2.0.
We use MockFog 2.0 to evaluate a fog-based smart factory application and showcase how its features can be used to study the impact of infrastructure changes and workload variations.
With these experiments, we also show that MockFog can achieve good experiment reproducibility, even in a public cloud environment.
\end{abstract}

\section{Introduction}\label{sec:introduction}

Fog computing is an emerging computing paradigm that promises to combine the benefits of edge computing and cloud computing~\cite{bermbach_research_2018,mahmud_fog_2018}.
For low latency, application components are deployed close to or near the edge, i.e., close to end users.
This can also reduce bandwidth consumption, mitigate privacy risks, and enable the edge to keep operating in the presence of network partitions.
For high scalability, application components can leverage stronger machines such as cloudlets~\cite{satyanarayanan_case_2009} within the core network~\cite{bermbach_research_2018} or run directly in the cloud.
This encompassing execution environment is commonly referred to as \textit{fog}~\cite{bonomi_fog_2012,bermbach_research_2018} and comprises all devices along the \enquote{cloud-to-thing continuum}\cite{openfog}.
However, even though fog computing has many advantages, there are currently only a few fog applications and \enquote{commercial deployments are yet to take off}~\cite{varshney_demystifying_2017}.
Arguably, the main adoption barrier is the deployment and management of physical infrastructure, particularly at the edge, which is in stark contrast to the ease of adoption in the cloud~\cite{bermbach_research_2018}.

In the lifecycle of a fog application, this is not only a problem when running and operating a production system -- it is also a challenge in application testing:
While basic design questions can be decided using simulation, e.g.,~\cite{brambilla_simulation_2014,hasenburg_supporting_2018,gupta_ifogsim:_2017}, there comes a point when a new application needs to be tested in practice.
The physical fog infrastructure, however, will typically be available for a very brief time only: in between having finished the physical deployment of devices and before going live.
Before that period, the infrastructure presumably does not exist and afterwards its full capacity is used in production.
Without an infrastructure to run more complex integration tests or benchmarks, e.g., for fault-tolerance in wide area deployments, however, the application developer is left with guesses, (very small) local testbeds, and simulation.
While approaches for the emulation of infrastructure testbeds exist, they typically focus on emulating edge devices, e.g.,~\cite{hashemian_wotbench:_2019,ramprasad_emu-iot_2019}.
Other approaches also emulate infrastructure within the core network or the cloud, but they miss support for automated experiment orchestration, e.g.,~\cite{coutinho_fogbed:_2018,mayer_emufog:_2017}.

In this paper, we extend our preliminary work presented in~\cite{hasenburg_mockfog:_2019}.
We propose to evaluate fog applications on an emulated infrastructure testbed created in the cloud which can be manipulated based on a pre-defined orchestration schedule.
In an emulated fog environment, virtual cloud machines are configured to closely mimic the real (or planned) fog infrastructure.
By using basic information on network characteristics, either obtained from the production environment or based on expectations and experiences with other applications, interconnections between the emulated fog machines can be manipulated to show similar characteristics.
Likewise, performance measurements from real fog machines can be used to determine resource limits on Dockerized\footnote{\url{https://docker.com}} application containers\footnote{When benchmarking Dockerized applications, developers need to account for Dockerization impacts~\cite{grambow_is_2019,grambow_dockerization_2018}.}.
This way, fog applications can be fully deployed in the cloud while experiencing comparable performance and failure characteristics as a real fog deployment.

Using an emulated infrastructure also makes it possible to change machine and network characteristics, as well as the workload used during application testing at runtime based on an orchestration schedule.
For example, this makes it possible to evaluate the impact of sudden machine failures or unreliable network connections as part of a system test with varying load.
While testing in an emulated fog will never be as \enquote{good} as in a real production fog environment, it is certainly better than simulation-based evaluation only.
Moreover, it allows application engineers to test arbitrary failure scenarios and various infrastructure options at large scale, which is also not possible on small local testbeds.

Thus, we make the following contributions:
\begin{itemize}
	\item We describe the design of MockFog, a system that can emulate fog computing infrastructure in arbitrary cloud environments, manage applications, and orchestrate experiments integrated into a typical application engineering process.
	\item We present our proof-of-concept implementation MockFog 2.0, the successor to the original proof-of-concept implementation\footnote{\url{https://github.com/OpenFogStack/MockFog-Meta}}.
	\item We demonstrate how MockFog 2.0 allows developers to automate experiments that involve changing infrastructure and workload characteristics with an example application.
\end{itemize}
The remainder of this paper is structured as follows: We first describe the design of MockFog and discuss how it is used within a typical application engineering process (\cref{sec:design}). Next, we evaluate our approach through a proof-of-concept implementation (\cref{sec:impl}) and a set of experiments with a smart-factory application using the prototype (\cref{sec:exp}).
Here, we also show that MockFog can achieve good experiment reproducibility, even in a public cloud environment.
Finally, we compare MockFog to related work (\cref{sec:related}) before a discussion (\cref{sec:discussion}) and conclusion (\cref{sec:conclusion}).

\section{MockFog Design} \label{sec:design}
In this section, we present the MockFog design, starting with a high level overview of its three modules (\cref{subsec:25_modules}).
Then, we discuss how to use Mockfog in a typical application engineering process (\cref{subsec:25_process}) before describing each of the modules (\crefrange{subsec:25_infra-emulation}{subsec:25_exp-orchestration}).

\subsection{MockFog Overview} \label{subsec:25_modules}

\begin{figure}[t]
    \centering
    \includegraphics[width=\columnwidth]{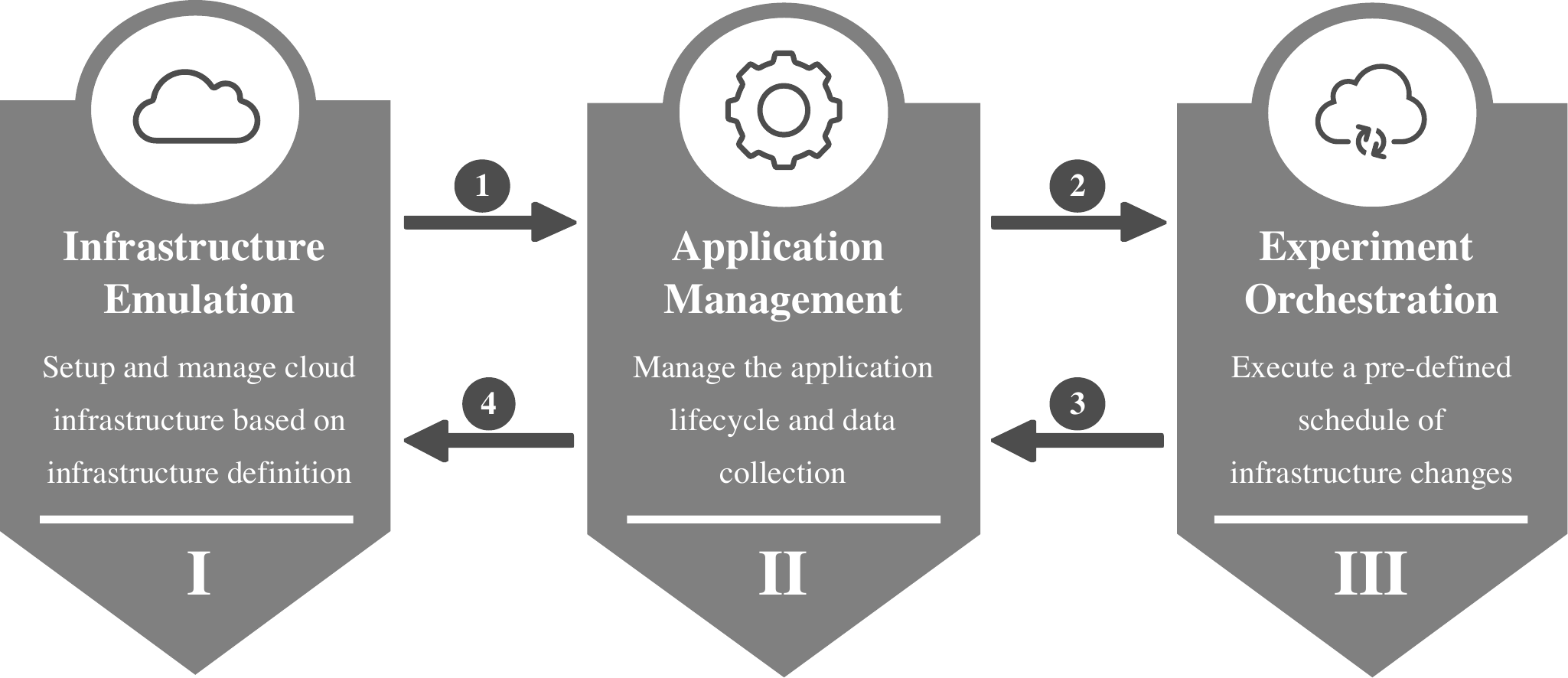}
    \caption{MockFog comprises three modules.}
    \label{fig:25_modules}
\end{figure}

MockFog comprises three modules: the infrastructure emulation module, the application management module, and the experiment orchestration module (see \cref{fig:25_modules}).
For the first module, developers model the properties of their desired (emulated) fog infrastructure, namely the number and kind of machines but also the properties of their interconnections.
The infrastructure emulation module uses this configuration for the infrastructure bootstrapping and infrastructure teardown.
For the second module, developers define application containers and where to deploy them.
The application management module uses this configuration for the application container deployment, the collection of results, and the application shutdown.
For the third module, developers define an experiment orchestration schedule that includes infrastructure changes and application instructions.
The experiment orchestration module uses this configuration to initiate infrastructure changes or to signal load generators\footnote{Requirements for load generators are highly application-specific and also depend on usage purposes such as benchmarking or testing. Since there is already a plethora of standard load generators, benchmarks, and application-specific ad-hoc load generators, we do not include a load generator in MockFog. Instead, we focus on integrating and managing arbitrary load generators through MockFog's signaling and orchestration features.} and the system under test at runtime.

The implementation of all three modules is spread over two main components: the \textit{node manager} and the \textit{node agents}.
There is only a single node manager instance in each MockFog setup.
It serves as the point of entry for application developers and is, in general, their only way of interacting with MockFog.
In contrast, one node agent instance runs on each of the cloud virtual machines (VMs) used to emulate the fog infrastructure.
Based on the node manager's input, node agents manipulate their respective VM to show the desired machine and network characteristics to the application.

\begin{figure}[t]
    \centering
    \includegraphics[width=\columnwidth]{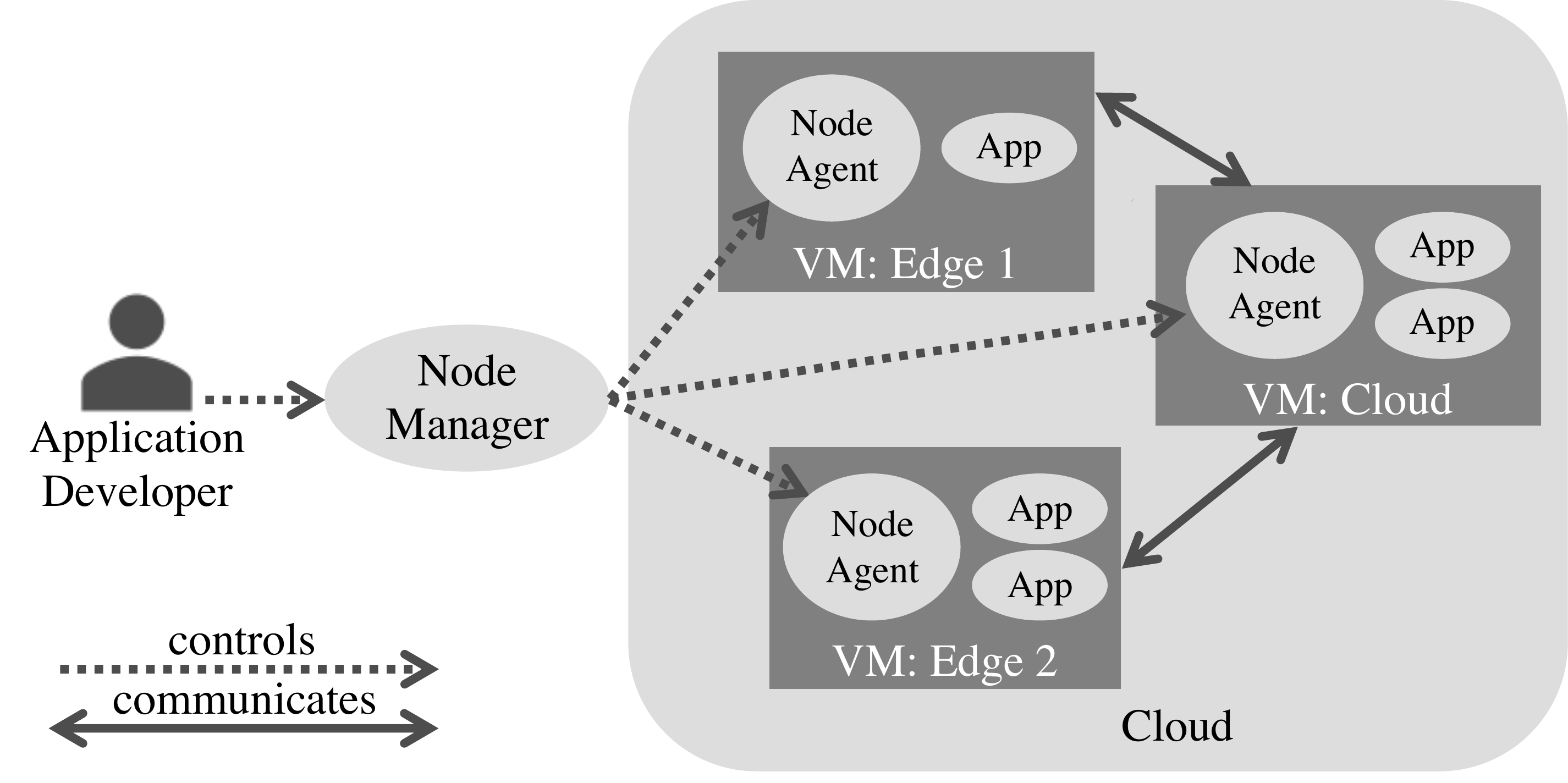}
    \caption{Example: MockFog node manager, node agents, and containerized application components (App).}
    \label{fig:25_nmanager_nagent}
\end{figure}

\Cref{fig:25_nmanager_nagent} shows an example with three VMs: two are emulated edge machines, and one is a single ``emulated'' cloud machine.
In the example, the node manager instructs the node agents to manipulate the network properties of their VMs in such a way that an application appears to have all its network traffic routed through the cloud VM.
Moreover, the node agents ensure that network manipulations do not affect communication to the node manager by using a dedicated management network.
Note that developers can freely choose where to run the node manager, e.g., it could run on a developer's laptop or on another cloud VM.

\subsection{Using MockFog in Application Engineering} \label{subsec:25_process}

A typical application engineering process starts with requirements elicitation, followed by design, implementation, testing, and finally maintenance.
In agile, continuous integration and DevOps processes, these steps are executed in short development cycles, often even in parallel -- with MockFog, we primarily target the testing phase.
Within the testing phase, a variety of tests could be run, e.g., unit tests, integration tests, system tests, or acceptance tests~\cite{winter_integrationstest_2013} but also benchmarks to better understand system quality levels of an application, e.g., performance, fault-tolerance, data consistency~\cite{bermbach_cloud_2017}.
Out of these tests, unit tests tend to evaluate small isolated features only, and acceptance tests are usually run on the production infrastructure; often, involving a gradual roll-out process with canary testing, A/B testing, and similar approaches, e.g.,~\cite{schermann_bifrost:_2016}.
For integration and system tests as well as benchmarking, however, a dedicated test infrastructure is required.
With MockFog, we provide such an infrastructure for experiments.

\begin{figure}[t]
    \centering
    \includegraphics[width=\columnwidth]{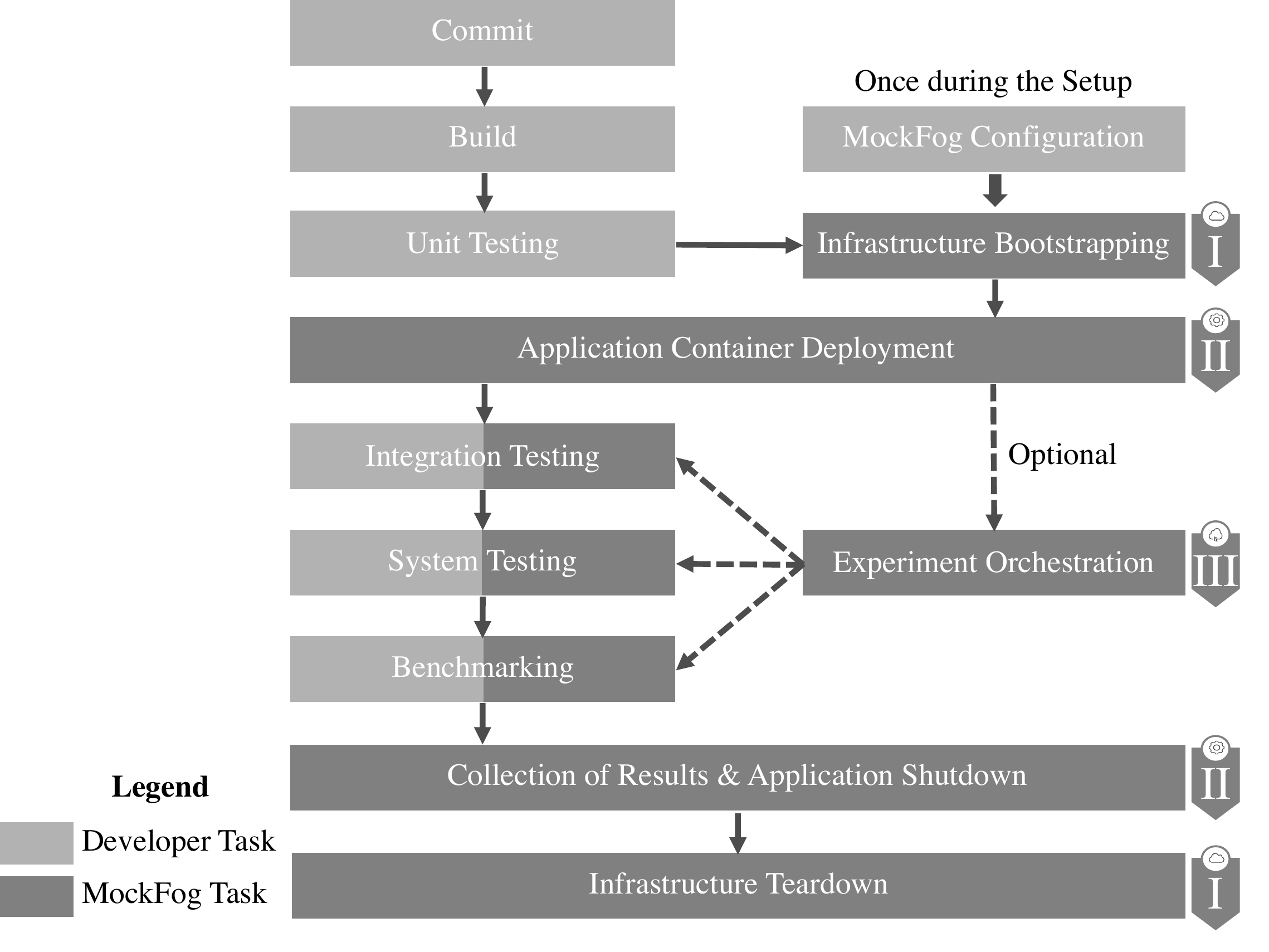}
    \caption{The three MockFog modules set up and manage experiments during the application engineering process.}
    \label{fig:25_application_engineering_process}
\end{figure}

We imagine that developers integrate MockFog into their deployment pipeline (see \cref{fig:25_application_engineering_process}) and use it with their existing continuous integration and deployment tooling.
Once a new version of the application has passed all unit tests, MockFog can be used to set up and manage experiments.
For the MockFog setup, a developer only needs to provide configuration files for the three MockFog modules, which we describe in more detail below.
We provide the configuration files used within our evaluation in \cref{sec:exp}.

\subsection{Infrastructure Emulation Module} \label{subsec:25_infra-emulation}

A typical fog infrastructure comprises several fog machines, i.e., edge machines, cloud machines, and possibly also machines within the network between edge and cloud~\cite{bermbach_research_2018}.
If no physical infrastructure exists yet, developers can follow guidelines, best practices or reference architectures such as proposed in~\cite{pfandzelter_iot_2019,santos_identifying_2019,karagiannis_comparison_2020,santos_architectural_2020,rausch_synthesizing_2020}.
On an abstract level, the infrastructure can be described as a graph comprising machines as vertices and the network between machines as edges~\cite{kajackas_internet_2010}.
In this graph, machines and network connections can also have properties such as the compute power of a machine or the available bandwidth of a connection.
For the infrastructure emulation module, the developer specifies such an abstract graph before assigning properties to vertices and edges.
We describe the machine and network properties supported by MockFog in \cref{sec:25_design_machines} and \cref{sec:25_design_connections}.

During the infrastructure bootstrapping step (see \cref{fig:25_application_engineering_process}), the node manager connects to the respective cloud service provider to set up a single VM in the cloud for each fog machine in the infrastructure model.
VM type selection is straightforward when the cloud service provider accepts the machine properties as input directly, e.g., on Google Compute Engine.
If not, e.g., on AWS EC2, the mapping selects the smallest VM that still fulfills the individual machine requirements.
MockFog then hides surplus resources by limiting resources for the containers directly.
When all machines have been set up, the node manager installs the node agent on each VM, which will later manipulate its VM's machine and network characteristics.

Once the infrastructure bootstrapping has been completed, the developer continues with the application management module.
Furthermore, MockFog provides IP addresses and access credentials for the emulated fog machines.
With these, the developer can establish direct SSH connections, use customized deployment tooling, or manage machines with the cloud service provider's APIs if needed.

Once all experiments have been completed, the developer can also use the infrastructure emulation module to destroy the provisioned experiment infrastructure.
Here, the node manager removes all emulated resources and deletes the access credentials created for the experiments.

\subsubsection{Machine Properties} \label{sec:25_design_machines}

Machines are the parts of the infrastructure on which application code is executed.
Fog machines can appear in various different flavors, ranging from small edge devices such as Raspberry Pis\footnote{\url{https://raspberrypi.org}}, over machines within a server rack, e.g., as part of a Cloudlet~\cite{satyanarayanan_case_2009,mahmud_fog_2018}, to virtual machines provisioned through a public cloud service such as AWS EC2. 

To emulate this variety of machines in the cloud, their properties need to be described precisely.
Typical properties of machines are compute power, memory, and storage.
Network I/O would be another standard property; however, we chose to model this only as part of the network in between machines.

While the memory and storage properties are self-explanatory, we would like to emphasize that there are different approaches to measuring compute power.
AWS EC2, for instance, uses the amount of vCPUs to indicate the compute power of a given machine.
This, or the number of cores, is a very rough approximation that, however, suffices for many use cases as typical fog application deployments rarely achieve 100\% CPU load.
It is also possible to use more generic performance indicators such as instructions per second (IPS) or floating-point operations per second (FLOPS).
Our current proof-of-concept prototype (\cref{sec:impl}) uses Docker's resource limits\footnote{\url{https://docs.docker.com/engine/reference/commandline/update/}}.

\subsubsection{Network Properties} \label{sec:25_design_connections}

Within the infrastructure graph, machines are connected through network connections: only connected machines can communicate.
In real deployments, these connections usually have diverse network characteristics~\cite{varshney_demystifying_2017}, e.g., slow and unreliable connections at the edge and fast and reliable connections near the cloud, which strongly affect applications running on top of them.
These characteristics, therefore, also need to be modeled -- see \cref{tbl:emu_connection_properties} for an overview of our model properties.
For example, if a connection between machines A and B has a delay of 10~ms, a dispersion of 2~ms, and a package loss probability of 5\%, a package sent from A to B would have a mean latency of 10~ms with a standard deviation of 2~ms and a 5\% probability of not arriving at all.

\begin{savenotes}
\begin{table}[t]
\renewcommand{\arraystretch}{1.3}
\caption{Properties of Emulated Network Connections}
\label{tbl:emu_connection_properties}
\centering
\begin{tabularx}{\columnwidth}{lX}\toprule
Property                    & Description                        \\ \midrule
Rate                        & Available Bandwidth Rate \\
Delay                       & Latency of Outgoing Packages       \\
Dispersion                  & Delay Dispersion (+/-)             \\
Loss                        & Percentage of Packages Lost in Transition \\
Corruption                  & Percentage of Corrupted Packages     \\
Reorder                     & Probability of Package Reordering        \\
Duplicate                   & Probability of Package Duplication    \\
\bottomrule
\end{tabularx}
\end{table}
\end{savenotes}

\begin{figure}[t]
    \centering
    \includegraphics[width=0.8\columnwidth]{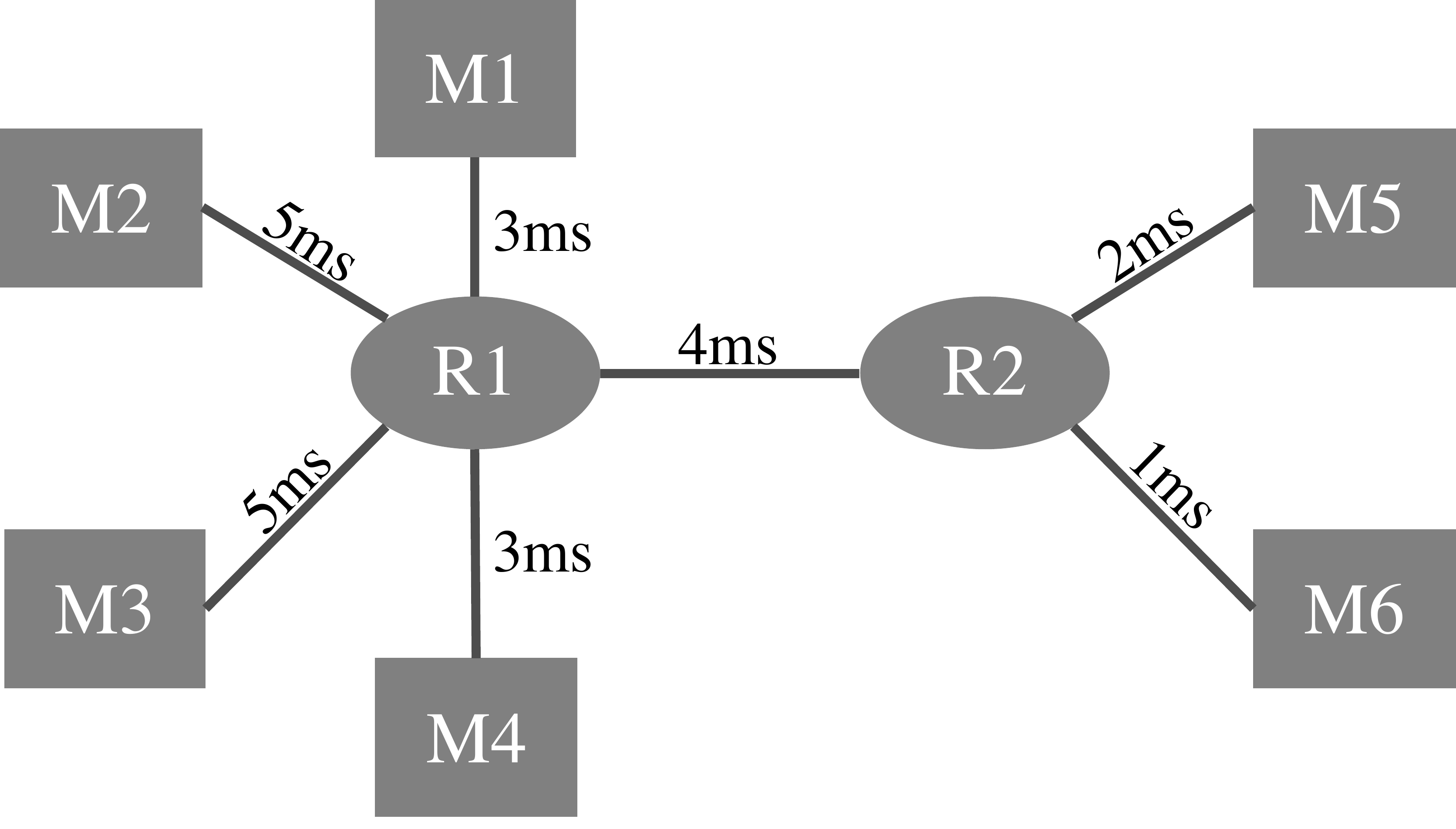}
    \caption{Example: Infrastructure graph with machines (M), routers (R), and network latency per connection.}
    \label{fig:25_graph_with_routers}
\end{figure}

In most scenarios, not all machines are connected directly to each other.
Instead, machines are connected via switches, routers, or other machines.
See \cref{fig:25_graph_with_routers} for an example with routers and imagine having to model the cartesian product of machines instead.
In the graph, network latency is calculated as the weighted shortest path between two machines.
For instance, if the connection between M2 and R1 (in short: M2-R1) has a delay of 5~ms, R1-R2 has 4~ms, and R2-M6 has 1~ms, the overall latency for M2-M6 is 10~ms.
The available bandwidth rate is the minimum rate of any connection on the shortest path between two machines.
The dispersion is the sum of dispersion values on the shortest path between two machines.
Probability-based metrics, e.g., loss, are aggregated along the shortest path between two machines using basic probability theory methods ($p= 1 - \prod_{i=1}^n (1-p_i)$).

\subsection{Application Management Module} \label{subsec:25_app-management}

Fog applications comprise many components with complex interdependencies.
The configuration of such an application also depends on the infrastructure as components are not deployed on a single machine.
For example, we need an IP address and port to communicate with a component running on another machine.
MockFog can deploy and configure application components on the emulated infrastructure, resolving such dependencies.
For this purpose, a requirement is that all application components are Dockerized.
Furthermore, developers have to define application containers and how they should be deployed on the infrastructure.
If these requirements cannot be met, developers can use their own deployment tooling instead of the application management module before continuing to the experiment orchestration.

\begin{figure}[t]
    \centering
    \includegraphics[width=0.9\columnwidth]{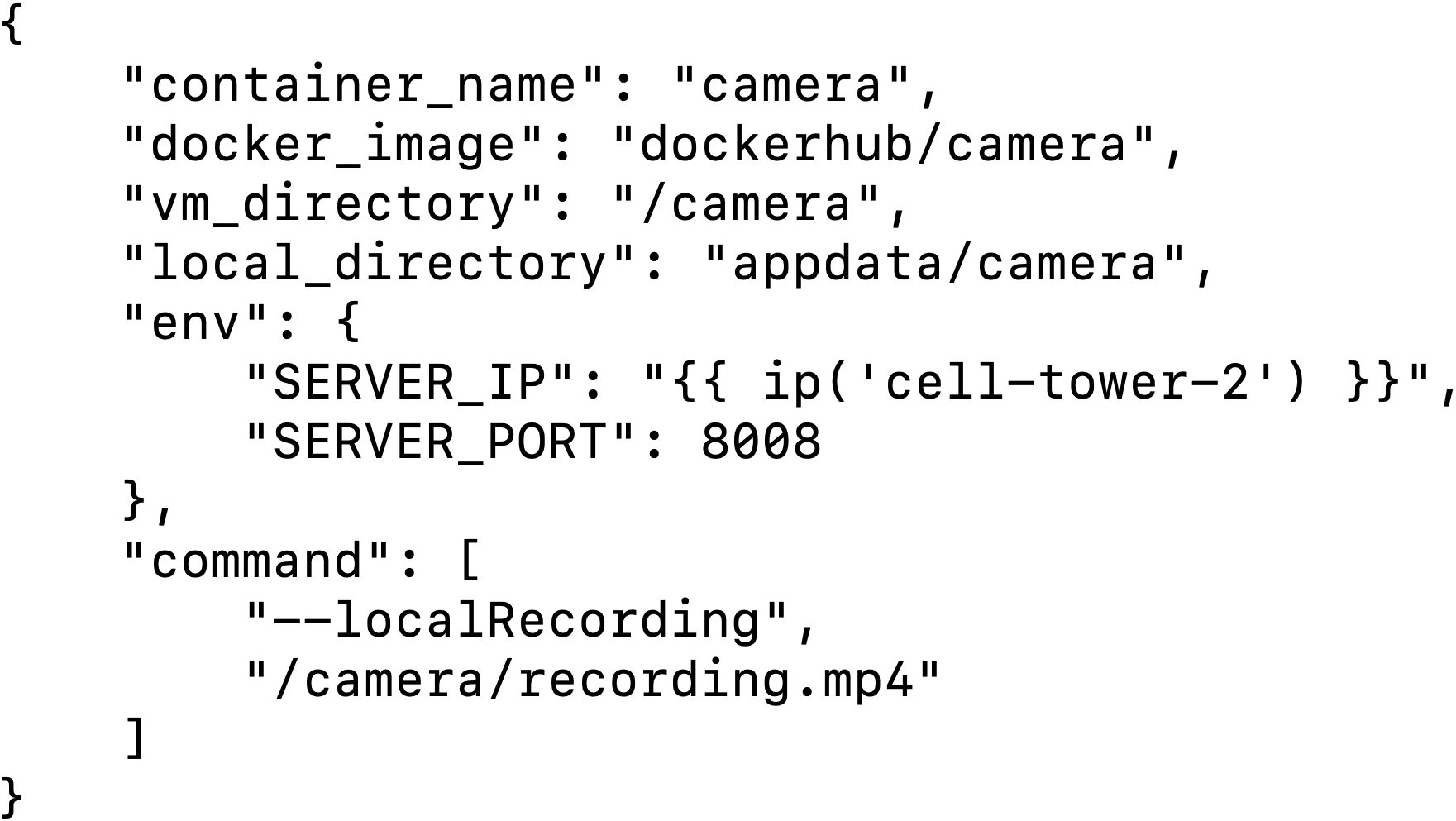}
    \caption{The container configuration comprises a unique container name and additional meta data.}
    \label{fig:25_camera}
\end{figure}

For each container, the container configuration specifies a unique container name, the Docker image to be used, information on local files that should be copied to the VM, environment variables, and command-line arguments.
As an example, \cref{fig:25_camera} shows a very simple container configuration for a container with name \textit{camera} in JSON format.
For this container, the Docker image is \textit{dockerhub/camera}; if it is not available locally, MockFog pulls the latest version from the Docker Hub.
Furthermore, MockFog copies the contents of the local directory \textit{appdata/camera} to the \textit{/camera} directory on each VM where the specific container will run.
When the container is started, the environment variables \textit{SERVER\_IP} and \textit{SERVER\_PORT} are set to the specified values and become available to the application running inside the container.
The value of \textit{SERVER\_IP} is resolved by a function that retrieves the IP address of the VM named \textit{cell-tower-2}.
Additional such functions, e.g., for retrieving the IP addresses of all VMs on which a container with a specific container name have been deployed, exist as well.
Finally, the \textit{camera} container is instructed to write a local copy of its recording to \textit{/camera/recording.mp4} via command-line arguments.
As this file path is inside the specified VM directory, its contents can be retrieved by MockFog automatically.

In the deployment configuration, developers specify for each container a deployment mapping of application components to VMs.
Furthermore, they can also limit CPU and memory resources available to a container, e.g., for balancing the resource needs of multiple containers running on the same VM.
During the application container deployment step (see \cref{fig:25_application_engineering_process}), the node manager installs dependencies on the VMs, copies files, and starts the configured containers.

Once the experiment has been completed, the developer can also use the application container module for terminating the application and for collecting results.

\subsection{Experiment Orchestration Module} \label{subsec:25_exp-orchestration}

There are various ways of testing and benchmarking an application.
As discussed in \cref{subsec:25_process}, MockFog primarily targets integration and system tests as well as benchmarking because these require a dedicated test infrastructure.
MockFog can artificially inject (and revert) failures to emulate network partitioning, simulate machine crashes and restarts, as well as other events for such experiments.
This is particularly useful as failures are common in real deployments but will not necessarily happen while an application is being tested.
Hence, artificial failures are the go-to approach for studying the fault-tolerance and resilience of an application~\cite{bermbach_towards_2014}.
While MockFog monitors the emulated infrastructure to detect deviations from what it configured, one might be interested in additional monitoring data.
For this purpose, we recommend to either use the tooling of the chosen cloud vendor, e.g., Amazon CloudWatch\footnote{\url{https://aws.amazon.com/cloudwatch}} when running on AWS, or to deploy custom tooling, e.g., Prometheus, alongside the application through the application management module.

For the experiment orchestration step (see \cref{fig:25_application_engineering_process}), developers define an orchestration schedule in the form of a state machine.
We describe the actions executed within a state in \cref{sec:25_state_lifecycle}; we describe how developers can build complex orchestration schedules with states and their transitions in \cref{sec:25_complex_schedules}.

\subsubsection{State Actions} \label{sec:25_state_lifecycle}

The orchestration schedule comprises a set of states and a set of transitioning conditions.
At each point in time, there is exactly one active state for which MockFog executes up to four actions in the following order (\cref{fig:25_state_lifecycle}):

\begin{figure}[t]
    \centering
    \includegraphics[width=\columnwidth]{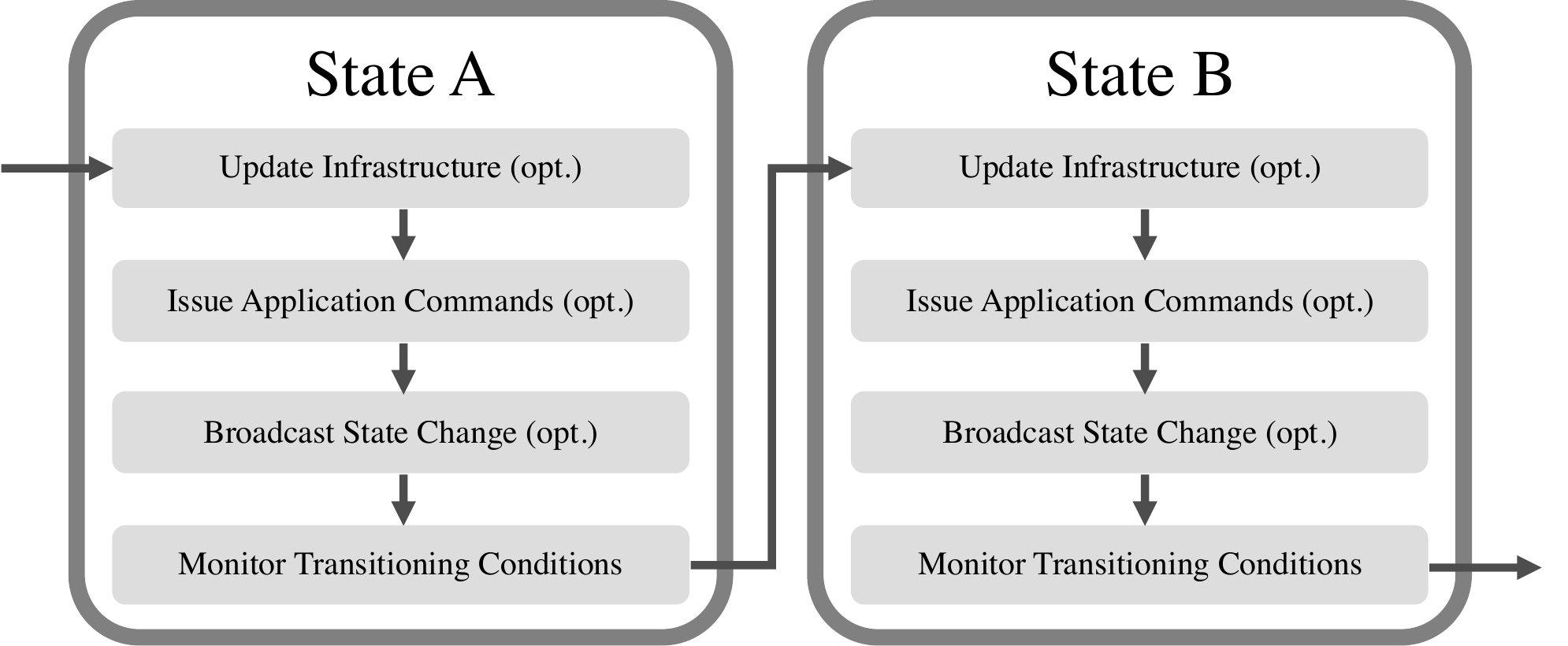}
    \caption{In each state, MockFog executes up to four actions.}
    \label{fig:25_state_lifecycle}
\end{figure}

\vspace{0.3cm}\noindent\textbf{Update Infrastructure (opt.)}\hspace{0.3cm}
With MockFog, all properties of emulated fog machines and network connections (\cref{tbl:emu_connection_properties}) can be manipulated.
For this, the node manager parses the orchestration schedule and sends instructions to the node agents which then update machine and network properties accordingly.
For example, it is possible to reduce the amount of available memory (e.g., due to noisy neighbors), render a set of network links temporarily unavailable, increase network latency or package loss, or render a machine completely unreachable, in which case the node agent blocks all (application) communication to and from the respective VM.
MockFog can also reset all infrastructure manipulations back to what was initially defined by the developer.
Node agents acknowledge infrastructure updates to assert adherence to the orchestration schedule.
If there are any problems that cannot be recovered autonomously, the node manager notifies the developer.
This action is optional.

\vspace{0.3cm}\noindent\textbf{Issue Application Commands (opt.)}\hspace{0.3cm}
Based on the orchestration schedule, the node manager can send customizable instructions to application components.
For example, this can be used to instruct a workload generator to change its workload profile.
This action is optional.

\vspace{0.3cm}\noindent\textbf{Broadcast State Change (opt.)}\hspace{0.3cm}
It is sometimes necessary to notify application components or a benchmarking system that a new state has been reached.
While the Issue Application Commands action may distribute complex scripts if necessary, this action is a lightweight notification mechanism.
In this, the first two actions are preparatory while this action signals to all components that the next experiment phase has been reached.
This action is optional.

\vspace{0.3cm}\noindent\textbf{Monitor Transitioning Conditions}\hspace{0.3cm}
Once the node manager reaches this action, an experiment timer is started.
The node manager then continuously monitors if a set of transitioning conditions -- as defined in the orchestration schedule -- have been met.
In MockFog, transitioning conditions can either be time-based or event-based:
A time-based condition is fulfilled when the experiment timer reaches the specified time threshold.
This is useful if a developer wishes to let the application run for a specific time to study effects of the active state.
An event-based condition is fulfilled when the node manager has received the required amount of a specific event (messages).
This is useful if a developer wishes to react to events distributed by application components, e.g., when any application component sends a \textit{failure} event once, MockFog should transition to the \textit{ABORT EXPERIMENTS} state.
If there are event-based conditions, application components have to either send events to the node manager directly or there must be a monitoring system such as Prometheus\footnote{\url{https://prometheus.io}} from which the node manager can receive events.

\subsubsection{Building Complex Orchestration Schedules} \label{sec:25_complex_schedules}

\begin{figure}[t]
    \centering
    \includegraphics[width=\columnwidth]{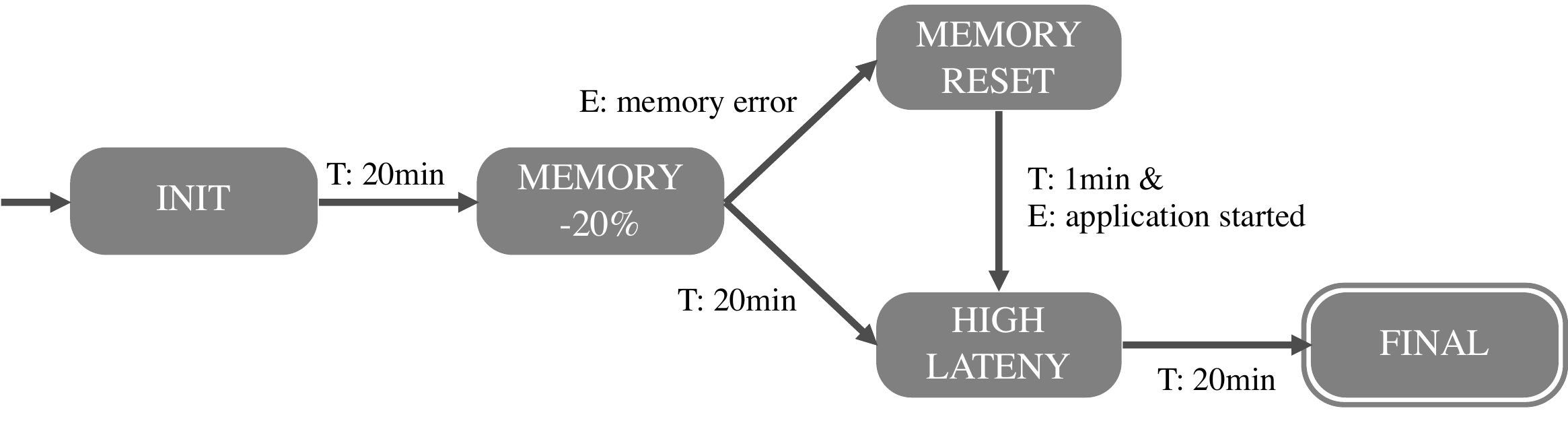}
    \caption{The experiment orchestration schedule can be visualized as a state diagram.}
    \label{fig:25_state-graph}
\end{figure}

For each state, developers can define multiple transitioning conditions; this allows MockFog to proceed to different states depending on what is happening during the experiment.
For instance, an orchestration schedule could have a time-based condition that leads to an \textit{ABORT EXPERIMENTS} state and additional event-based conditions that lead to a \textit{NEXT LOAD PHASE} state.
A transitioning condition may comprise several sub-conditions connected by boolean operators.
This allows developers to define arbitrarily complex state diagrams, see for example \cref{fig:25_state-graph}.

In the example, the orchestration schedule comprises five states; the arrows between states resemble the transitioning conditions.
When started, the node manager transitions to \textit{INIT}, i.e., it distributes the infrastructure configuration update and application commands.
Afterward, it broadcasts state change messages (e.g., this might initiate the workload generation needed for benchmarking) and begins monitoring the transition conditions of \textit{INIT}.
As the only transitioning condition is a time-based condition set to 20 minutes (T: 20min), the node manager transitions to \textit{MEMORY -20\%} once it has been in the \textit{INIT} state for 20 minutes.
During \textit{MEMORY -20\%}, the node manager instructs all node agents to reduce the amount of memory available to application components by 20\% via the Update Infrastructure action.
Then it again broadcasts state change messages (e.g., this might restart workload generation) and starts to monitor the transitioning conditions of \textit{MEMORY -20\%}.
For this state, there are two transitioning conditions.
If any application component emits a \textit{memory error} event, the node manager immediately transitions to \textit{MEMORY RESET} and instructs the node agents to reset memory limits.
Otherwise, the node manager transitions to \textit{HIGH LATENCY} after 20 minutes.
It also transitions to \textit{HIGH LATENCY} from \textit{MEMORY RESET} when it receives the event \textit{application started} and at least one minute has elapsed.
At the start of \textit{HIGH LATENCY}, the node manager instructs all node agents to increase the latency between emulated machines.
Then, it again broadcasts state change messages and waits for 20 minutes before finally transitioning to \textit{FINAL}.

\section{Proof-of-Concept Implementation} \label{sec:impl}

In this section, we describe our proof-of-concept implementation MockFog 2.0.
Due to the significant changes to the MockFog approach, MockFog 2.0 is a complete rewrite and does not build on the first MockFog prototype.
MockFog 2.0 has been developed with the goal of independence from specific IaaS cloud providers and can therefore be extended to the provider of choice.
Our current open source proof-of-concept prototype\footnote{\url{https://github.com/OpenFogStack/MockFog2}} integrates with AWS EC2.
By using EC2, MockFog has the same benefits and disadvantages as this cloud service:
acquiring machines is easy and inexpensive, but experiments might be affected by factors such as busy neighbors.
Thus, if higher emulation accuracy is needed, MockFog can easily be extended to also support other cloud services such as Grid'5000\footnote{\url{https://www.grid5000.fr/}} or bare-metal machines.
For this, it is sufficient to add another Ansible playbook to the infrastructure module.
Our implementation contains two NodeJS packages: the node manager (\cref{subsec:25_node-manager}) and the node agent (\cref{subsec:25_node-agent}).

\subsection{Node Manager} \label{subsec:25_node-manager}

The node manager NodeJS package can either be integrated with custom tooling or be controlled via the command-line.
We provide a command-line tool as part of the package that allows users to control the three modules' functionality.
For the infrastructure emulation module, the node manager relies on the Infrastructure as Code (IaC) paradigm.
Following this paradigm, an infrastructure definition tool serves to ``define, implement, and update IT infrastructure architecture''~\cite{morris_infrastructure_2016}.
The main advantage of this is that users can define infrastructure in a declarative way with the IaC tooling handling resource provisioning and deployment idempotently.
In our implementation, the node manager relies on Ansible\footnote{\url{https://ansible.com}} playbooks.

The node manager command-line tool offers several commands for each module.
As part of the infrastructure emulation module, the developer can:
\begin{itemize}
	\item Bootstrap machines: set up virtual machines on AWS EC2 and configure a virtual private cloud and the necessary subnets.
	\item Install node agents: (re)-install the node agent on each VM.
	\item Modify network characteristics: instruct node agents to modify network characteristics.
	\item Destroy and clean up: remove all resources and delete everything created through the \textit{bootstrap machines} command.
\end{itemize}

When modifying the network characteristics for a MockFog-deployed application, the node manager accounts for the latency between provisioned VMs.
For example, when communication should, on average, incur a 10~ms latency, and the existing average latency between two VMs is already 0.7~ms, the node manager instructs the respective node agents to delay messages by 9.3~ms.

\noindent As part of the application management module, the developer can:
\begin{itemize}
	\item Prepare files: upload the local application directories to the VMs and pull Docker images.
	\item Start containers: start Docker containers on each VM and apply container resource limits.
	\item Stop containers: stop Docker containers on each VM.
	\item Collect results: download the application directories from the VMs to a local directory on the node manager machine.
\end{itemize}

With the experiment orchestration module, the developer can initialize experiment orchestration.
When the orchestration schedule includes infrastructure changes, the node manager instructs affected node agents to override their current configuration following the updated model.
This is done via a dedicated ``management network'', which always has vanilla network characteristics and is hidden from application components.

\subsection{Node Agent} \label{subsec:25_node-agent}

While the node agent is also implemented in NodeJS, it uses the Python library \texttt{tcconfig}\footnote{\url{https://github.com/thombashi/tcconfig}} to manage network connections.
\textsc{tcconfig} is a command wrapper for the linux traffic control utility \texttt{tc}\footnote{\url{https://man7.org/linux/man-pages/man8/tc.8.html}}.
Thus, our current node agent prototype only works for Linux-based VMs.
The node manager ensures that all dependencies are installed alongside the node agent.

The node agent can either be started by the node manager or manually via command-line.
The only configuration necessary is the port on which the node agent exposes its REST endpoint.
This REST endpoint is used by the node manager but can also be used by developers directly.
To simplify its usage, we created a fully documented Swagger\footnote{\url{https://swagger.io/}} interface.

Using the REST endpoint, one can retrieve status information and real-time \texttt{ping} measurements to a list of other machines.
The node manager uses the \texttt{ping} measurement results to calculate the artificial delay, which should be injected to reach the desired latency between VMs.
Furthermore, the endpoint can be used to set resource limits for individual containers as needed by the application management module for the \textit{start containers} command and by the experiment orchestration module.
Finally, the endpoint can be used to supply (and read the current) network manipulation configuration.
On each update call, the node agent receives an adjacency list containing all other VMs.
The list includes the corresponding specification of its effective metrics: how it should be realized from the viewpoint of the node manager's infrastructure model.
If a particular machine should not be reachable, the adjacency list contains a package loss probability of 100\% for the corresponding VM.
This allows us to emulate network partitions easily.

\section{Experiments} \label{sec:exp}
After having shown with our proof-of-concept implementation MockFog 2.0 that this approach can indeed be implemented, we now use an example application to showcase its key features.
In this second evaluation part, we run experiments with a fog-based smart factory application (\cref{sec:35_app-overview}) for which we emulate a runtime infrastructure with MockFog 2.0.
In the experiments, we use an orchestration schedule (\cref{sec:35_orchestration-schedule}) that includes multiple infrastructure and workload changes and study the effects on the application (\cref{sec:35_results}).
Note that our goal is to provide an overview of the features of MockFog 2.0 and not to design a realistic benchmark or system test of our example application.
This evaluation also serves the purpose of showing how simple it is to run such application experiments with MockFog: one only needs to create configuration files\footnote{The configuration files of our experiments are available in the \textit{node-manager/run-example-smartfactory} directory of our code repository.} for our three modules.
Then, MockFog sets up an infrastructure testbed in the cloud, handles the application roll-out and experiment orchestration, collects results (i.e., application output such as log files and the log files of node agents), and finally destroys the testbed.
As a scenario, we build upon the smart factory application introduced in~\cite{pf2020zero}\footnote{Our application source code is available at \url{https://github.com/OpenFogStack/smart-factory-fog-example/tree/mockfog2}.}.

\subsection{Overview of the Smart Factory Example} \label{sec:35_app-overview}

In the smart factory, a production machine produces goods that are packaged by another machine.
Based on input from a camera and a temperature sensor, the production rate and packaging rate are adjusted in real-time.
Furthermore, the packaging rate is used to create a logistic prognosis, i.e., for scheduling the collection and delivery of goods.
Finally, a dashboard provides a historic packaging rate overview.

\begin{figure}[ht!]
    \centering
    \includegraphics[width=\columnwidth]{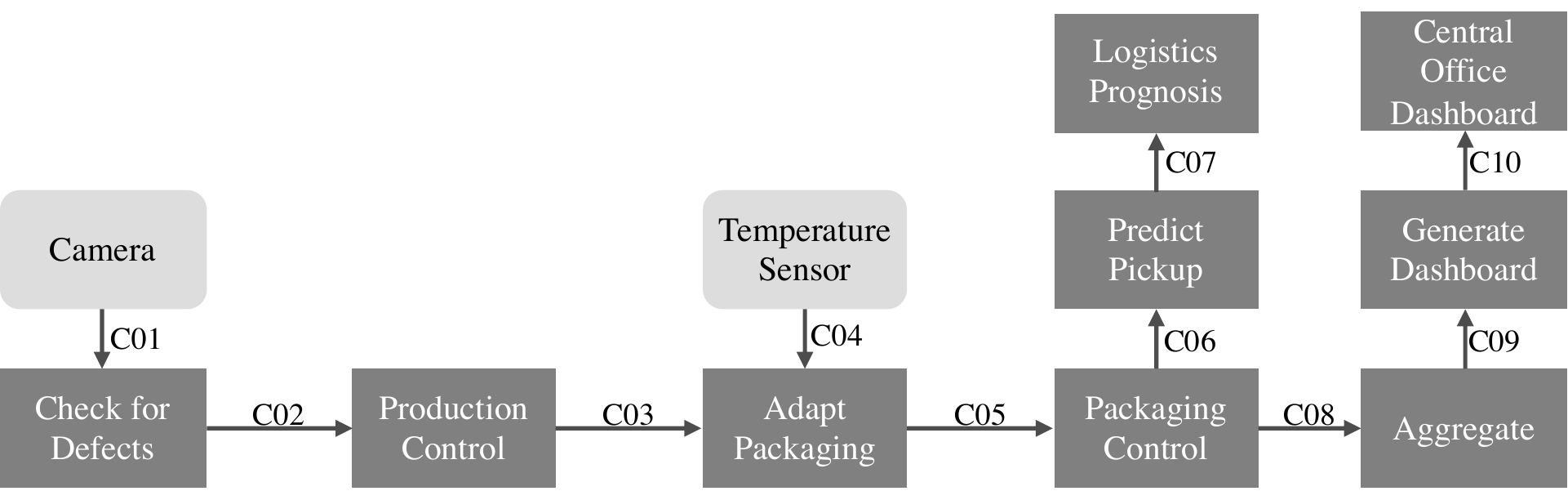}
    \caption{The smart factory application comprises 11 components and 10 communication paths between individual components (C01 --- C10).}
    \label{fig:35_application}
\end{figure}

Each of the components of the smart factory application communicates with at least one other component (see \cref{fig:35_application}).
\textit{Camera} sends its recordings to \textit{check for defects} which notifies \textit{production control} about products that should be discarded.
Based on input from \textit{production control} and \textit{temperature sensor}, \textit{adapt packaging} transmits the target packaging rate to \textit{packaging control}.
\textit{Adapt packaging} calculates the packaging rate based on the current production rate, the backlog of produced but not packaged items, and the temperature input:
packaging must be halted if the current temperature exceeds a threshold.
\textit{Packaging control} sends the current rate and backlog to \textit{predict pickup} and \textit{aggregate}.
\textit{Predict pickup} predicts when the next batch of goods is ready for pickup and sends this information to \textit{logistics prognosis}.
\textit{Aggregate} aggregates multiple rate and backlog values to preserve bandwidth and transmits the results to \textit{generate dashboard}.
\textit{Generate dashboard} stores the data in a database, creates an executive summary, and sends it to \textit{central office dashboard}.

The smart factory application comprises components that react to events from the physical world (light gray boxes) and components that only react to messages received from other application components (dark gray boxes).
For example, the \textit{temperature sensor} measures the physical machine's operation temperature that packages goods.
\textit{Adapt packaging}, on the other hand, receives messages from other application components and has no direct interaction with the physical world.

When testing real-time systems, two important concepts are reproducibility and controllability~\cite[p. 263]{ammann_introduction_2008}.
During experiments, \textit{camera} and \textit{temperature sensor} hence generate an input sequence that can be controlled by MockFog 2.0 to achieve reproducibility.
Furthermore, components that do something in the physical world based on received messages, e.g., \textit{packaging control}, only log their actions when doing experiments rather than sending instructions to physical machines.

\begin{figure}[ht!]
    \centering
    \includegraphics[width=\columnwidth]{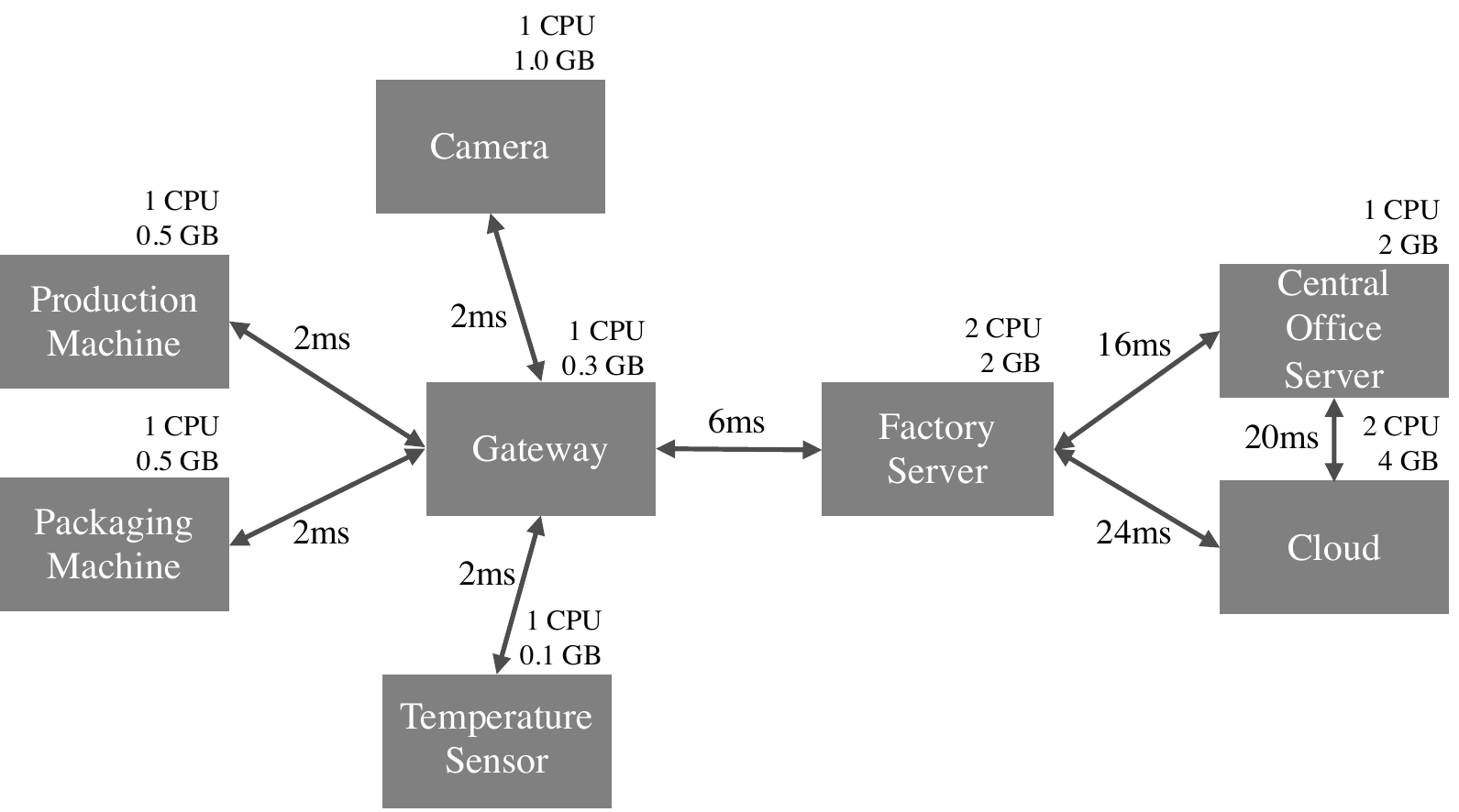}
    \caption{The smart factory infrastructure comprises multiple machines with different CPU and memory resources. Communication between directly connected machines incurs a round-trip latency between 2~ms and 24~ms.}
    \label{fig:35_infrastructure}
\end{figure}

\begin{table}[ht!]
\renewcommand{\arraystretch}{1.3}
\caption{Mapping of Application Components to Machines}
\label{tbl:deployment_mapping}
\centering
\begin{tabularx}{\columnwidth}{lX}\toprule
Application Component       & Machine \\ \midrule
Camera                      & Camera \\
Temperature Sensor          & Temperature Sensor \\
Check for Defects           & Gateway \\
Adapt Packaging             & Gateway \\
Production Control          & Production Machine \\
Packaging Control           & Packaging Machine \\
Predict Pickup              & Factory Server \\
Logistics Prognosis         & Factory Server \\
Aggregate                   & Factory Server \\
Generate Dashboard          & Cloud \\
Central Office Dashboard    & Central Office Server \\
\bottomrule
\end{tabularx}
\end{table}

\Cref{fig:35_infrastructure} shows the machines and network links of the smart factory infrastructure.
A gateway connects the camera, production machine, packaging machine, and temperature sensor.
Each has a 2~ms round-trip latency to the gateway, 1 CPU core, and 0.1~GB to 1.0~GB of available memory.
The gateway is connected to the factory server, which is connected to the central office server and the cloud.
The cloud and central office server are also connected directly.
All connections between machines are set up bandwidth of 1~GBit and do not incur any package loss, corruption, reordering, or duplicates.
\Cref{tbl:deployment_mapping} shows the mapping of application components to machines.
To derive such a mapping and to compare it to other approaches, developers can use approaches such as~\cite{skarlat_optimized_2017,hasenburg_supporting_2018,khare_linearize_2019}.

\subsection{Orchestration Schedule} \label{sec:35_orchestration-schedule}

\begin{figure}[t]
    \centering
    \includegraphics[width=0.9\columnwidth]{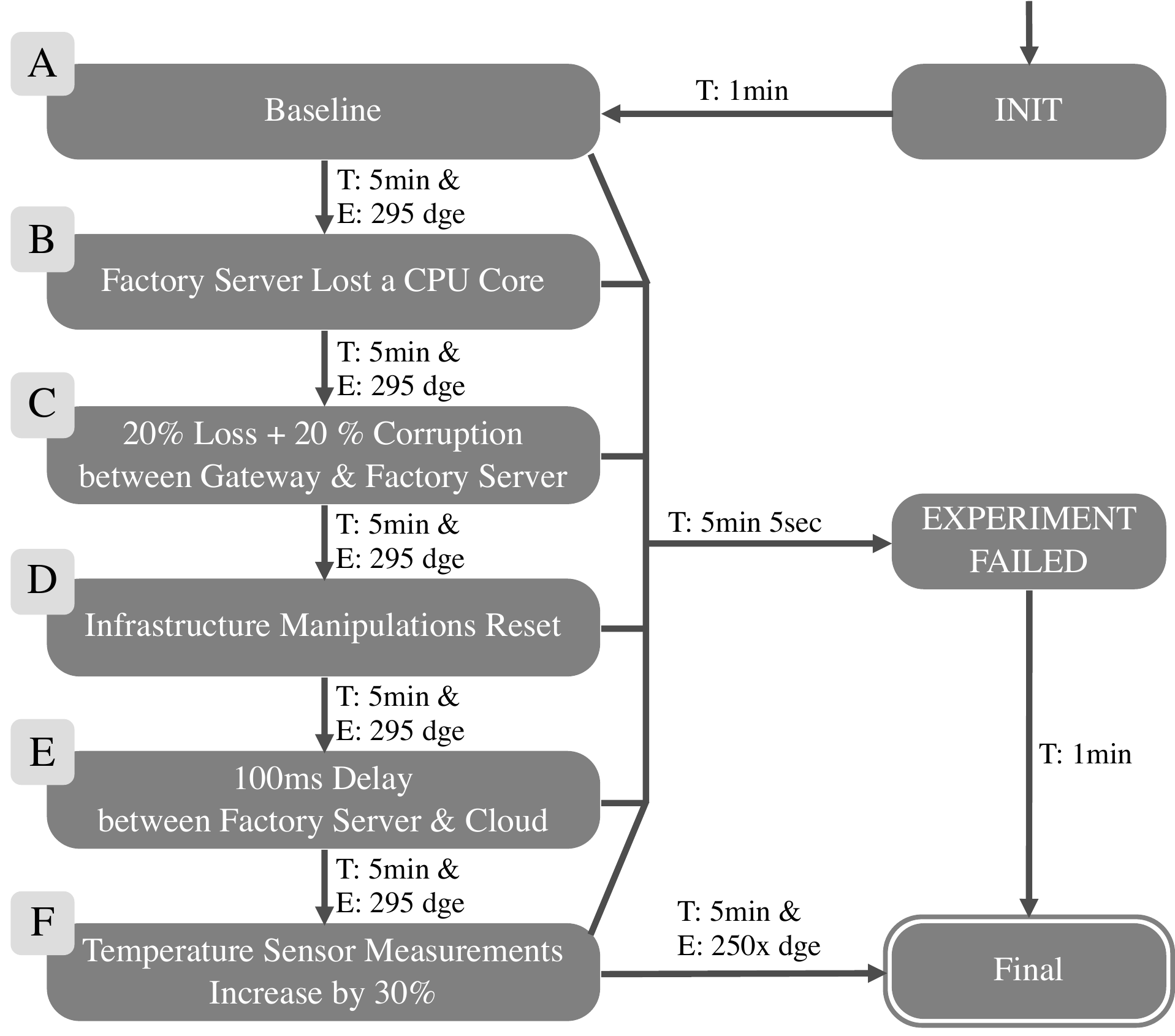}
    \caption{The orchestration schedule has nine states. During successful executions, the transitioning conditions mostly use a combination of time-based (5 minutes) and event-based conditions (receipt of 295 dashboard generated events (dge)).}
    \label{fig:35_orchestration_schedule}
\end{figure}

\begin{figure*}[t]
    \centering
    \includegraphics[width=\textwidth]{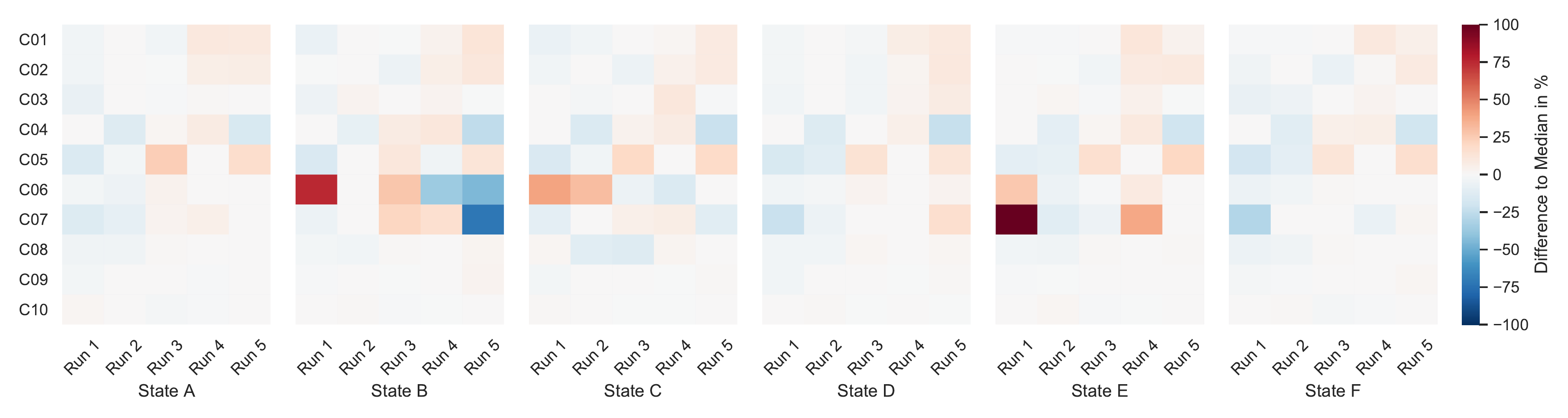}
    \caption{Latency deviation across experiment runs is small for most communication paths even though experiments were run in the cloud. On paths C06 and C07, resource utilization is high in states \textit{B}, \textit{C}, and \textit{E} leading to the expected variance across experiment runs.}
    \label{fig:35_reproducibility_latency}
\end{figure*}

For the experiments, we use an orchestration schedule with nine states (\cref{fig:35_orchestration_schedule}).
At the beginning of each state, MockFog 2.0 instructs \textit{camera} and \textit{temperature sensor} to restart their workload data sequence.
Thus, the application workload is comparable during each state.
The schedule starts with \textit{INIT}; after a minute, MockFog 2.0 transitions to state \textit{A}.
The purpose of state \textit{A} is to establish a baseline by running the application in an environment that closely mimics the real production environment.
At runtime, \textit{generate dashboard} creates a new dashboard once per second and sends a notification to the node manager.
We use this event as a failure indicator in all states; if it has not been received at least 295 times within a timeframe of 5 minutes, the experiment failed.
If, however, it has been received 295 times and five minutes have passed, MockFog 2.0 transitions to the next state.

For state \textit{B}, MockFog 2.0 changes the infrastructure:
the factory server has only access to one CPU core instead of two.
Then, in state \textit{C}, loss and corruption on the network link between gateway and factory server are set to 20\%.
Note that the factory server has not regained access to its second CPU core.
In state \textit{D}, all infrastructure changes are reset; the environment now again closely mimics the real production environment and the application can stabilize.
In state \textit{E}, the round-trip latency for messages sent from the factory server to the cloud is increased from 24~ms to 100~ms.
In state \textit{F}, the latency is reset to 24~ms but the temperature sensor is instructed to change the measured generation:
the average temperature sensor measurements are now 30\% higher which causes the packaging machine to pause more frequently.
This, in turn, should decrease the average packaging rate and increase the average packaging backlog.
After state \textit{F}, MockFog 2.0 transitions to \textit{FINAL} and the experiment orchestration ends.
If at any point a failure occurs, MockFog 2.0 will transition to \textit{EXPERIMENT FAILED}.

\subsection{Results} \label{sec:35_results}

In the following, we first validate that running the orchestration schedule leads to reproducible results (\cref{sec:35_validate_reproducibility}).
Then, we analyze how the changes made in each state of the orchestration schedule affect the smart factory application (\cref{sec:35_state_analysis}) and summarize our results (\cref{sec:35_summary}).
Note that the analysis of application logs and other output files is not done by MockFog since this is entirely application-specific and also depends on the load generator used.
Interpreting which latency, processing time, respective variance values, etc. are acceptable, also depends on the concrete application.
Thus, our results and conclusions are only valid for our specific use case and primarily serve the purpose of demonstrating how this process could look like in practice.

\subsubsection{Experiment Reproducibility} \label{sec:35_validate_reproducibility}

To analyze reproducibility, we repeat the experiment five times.
For each experiment run, we bootstrap a new infrastructure, install the application containers, and start the experiment orchestration -- this is done automatically by MockFog 2.0.
After the experiment run, we calculate the average (one-way) latency for each communication path (C01 to C10 in \cref{fig:35_application}).
Ideally, the latency results from all five runs should be identical for each communication path; in the following, we refer to the five measurement values for a given communication path as latency set.
In practice, however, it is not possible to achieve such a level of reproducibility because the application is influenced by outside factors~\cite[p. 263]{ammann_introduction_2008}.
For example, running an application on cloud VMs and in Docker containers already leads to significant performance variation~\cite{grambow_is_2019,grambow_dockerization_2018}.
To measure this variation, we use the median runs of each latency set as a baseline and calculate how much individual runs deviate from this baseline (see \cref{fig:35_reproducibility_latency}).
Considering that we are running the experiments in a public cloud, we can see in the figure that the deviation is small for almost all communication paths.
The exception are communication paths C06 and C07 in states \textit{B}, \textit{C}, and \textit{E} which show significant variance across runs.
In these states, the node manager applies various resource limits on the factory server.
Reducing the available compute and network resources seems to impact the stability of affected communication paths negatively.
Identifying such cases, however, in which infrastructure changes negatively impact application stability is exactly for what we designed MockFog.
Thus, we can conclude that experiment orchestration leads to reproducible results under normal operating conditions.
This holds true even if a new set of virtual machines is allocated for each run.
When a higher experiment reproducibility is required than the one achievable in a public cloud, one could, for example, add support for private OpenStack clusters to the infrastructure module.

\subsubsection{Application Impact of State Changes} \label{sec:35_state_analysis}

Of the five experiment runs, the second run is the most representative for the orchestration schedule:
On average, the latency of its communication paths deviate by 4.45\% from the median latency of the set.
Run three, four, five, and one deviate by 5.20\%, 5.28\%, 9.77\% and 10.23\% respectively.
Thus, we select the second run as the basis for analyzing how the changes made in each state affect application metrics.

\begin{figure}[t]
    \centering
    \includegraphics[width=\columnwidth]{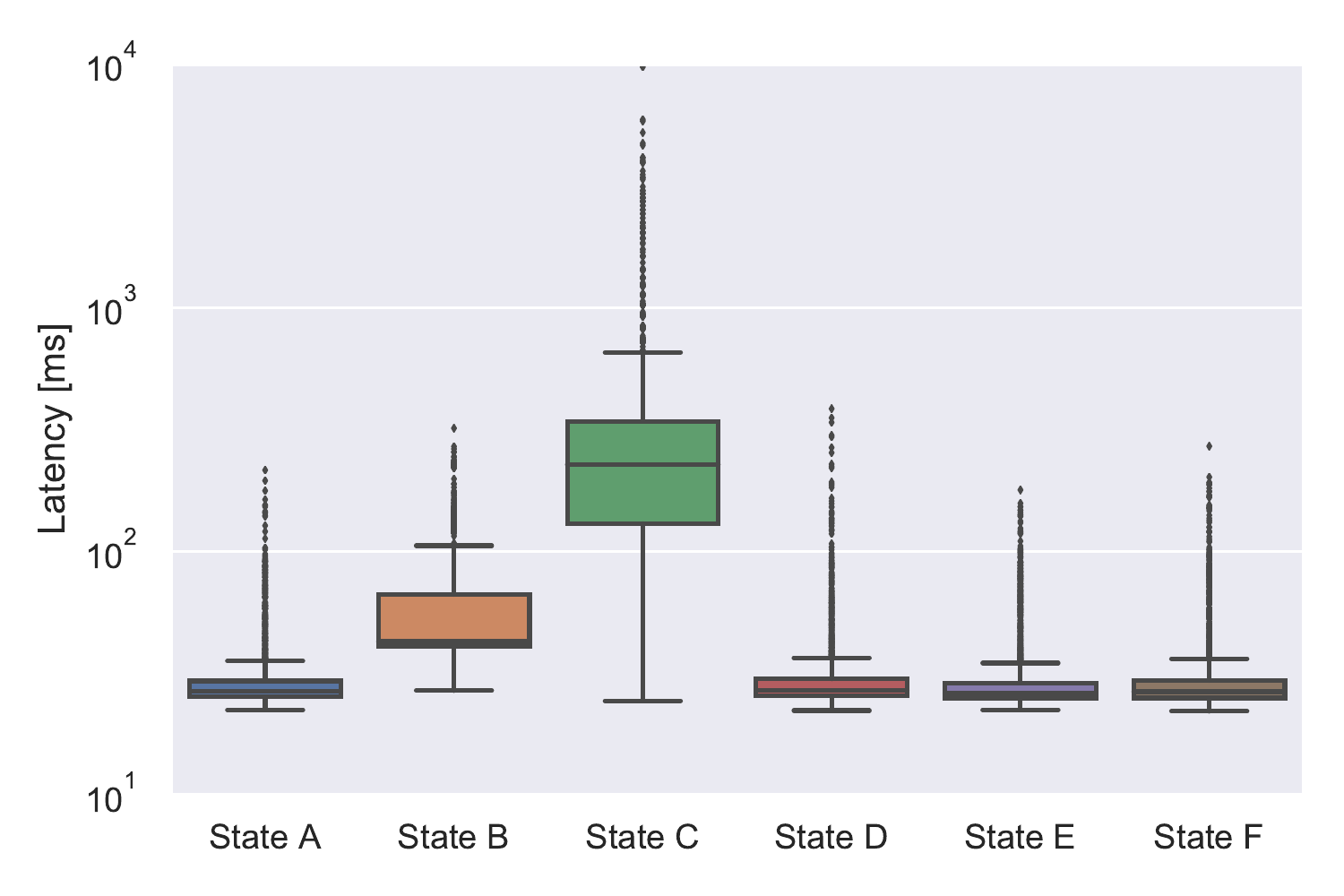}
    \caption{Latency between \textit{packaging control} and \textit{logistics prognosis} is affected by both CPU and network restrictions.}
    \label{fig:35_packaging-logistics}
\end{figure}
\Cref{fig:35_packaging-logistics} shows the latency between \textit{packaging control} and \textit{logistics prognosis}.
This latency includes the communication path latency of C06 and C07, as well as the time \textit{predict pickup} needs to create the prognosis.
In states \textit{A}, \textit{D}, \textit{E}, and \textit{F}, there are either no infrastructure changes or the ones made are on alternative communication paths; thus, latency is almost identical.
In state \textit{B}, the factory server loses a CPU core; as a result, \textit{predict pickup} needs more time to create a prognosis which increases the latency.
In state \textit{C}, the communication path C06 additionally suffers from a 20\% probability of package loss and a 20\% probability of package corruption.
As these packages have to be resent\footnote{Resent packages can also be impacted by loss or corruption.}, this significantly increases overall latency.

\begin{figure}[t]
    \centering
    \includegraphics[width=\columnwidth]{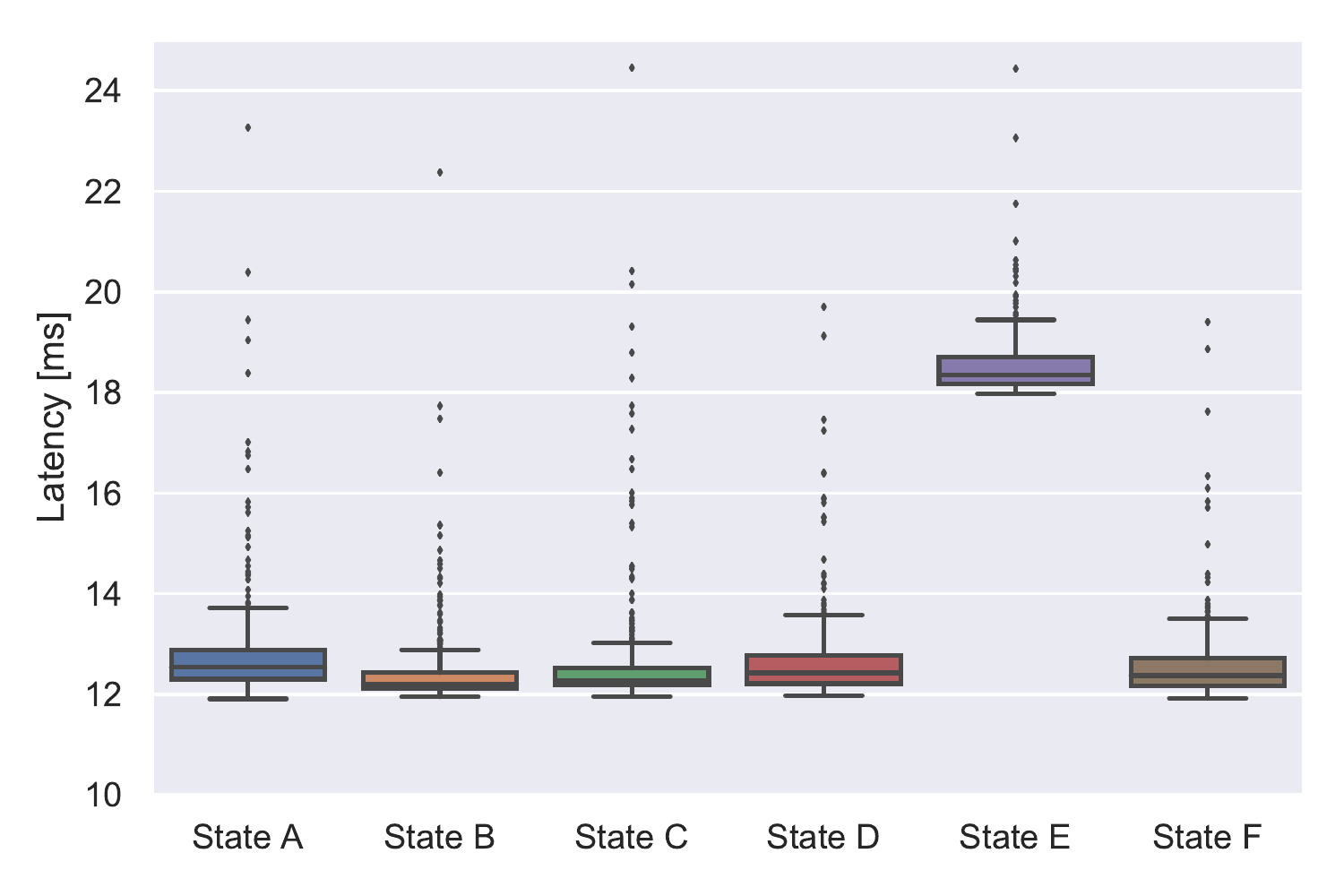}
    \caption{Latency on C09 between \textit{aggregate} and \textit{generate dashboard} is affected by the delay between factory server and cloud.}
    \label{fig:35_aggregate-generate}
\end{figure}
\Cref{fig:35_aggregate-generate} shows the latency on C09, i.e., the time between \textit{aggregate} sending and \textit{generate dashboard} receiving a message.
In states \textit{A}, \textit{D}, and \textit{F}, there are either no infrastructure changes or the ones made are on alternative communication paths; thus, latency is almost identical.
Note that the minimum latency is 12~ms; this makes sense as the round-trip latency between factory server and cloud is 24~ms.
In states \textit{B} and \textit{C}, the factory server loses a CPU core; MockFog 2.0 implements this limitation by setting Docker resource limits.
As a result, there is now 1 CPU core that is not used by the application containers and hence available to the operating system.
As the resource limitation seems not to impact \textit{aggregate}, the additional operating system resources slightly decrease latency.
While the effect here is only marginal, one has to keep such side effects in mind when doing experiments with Docker containers.
In state E, the round-trip latency between factory server and cloud is increased to 100~ms.
Still, the minimum (one-way) latency only increases to 18~ms as packages are routed via the central office server (round-trip latency is 16~ms + 20~ms).

\begin{figure}[t]
    \centering
    \includegraphics[width=\columnwidth]{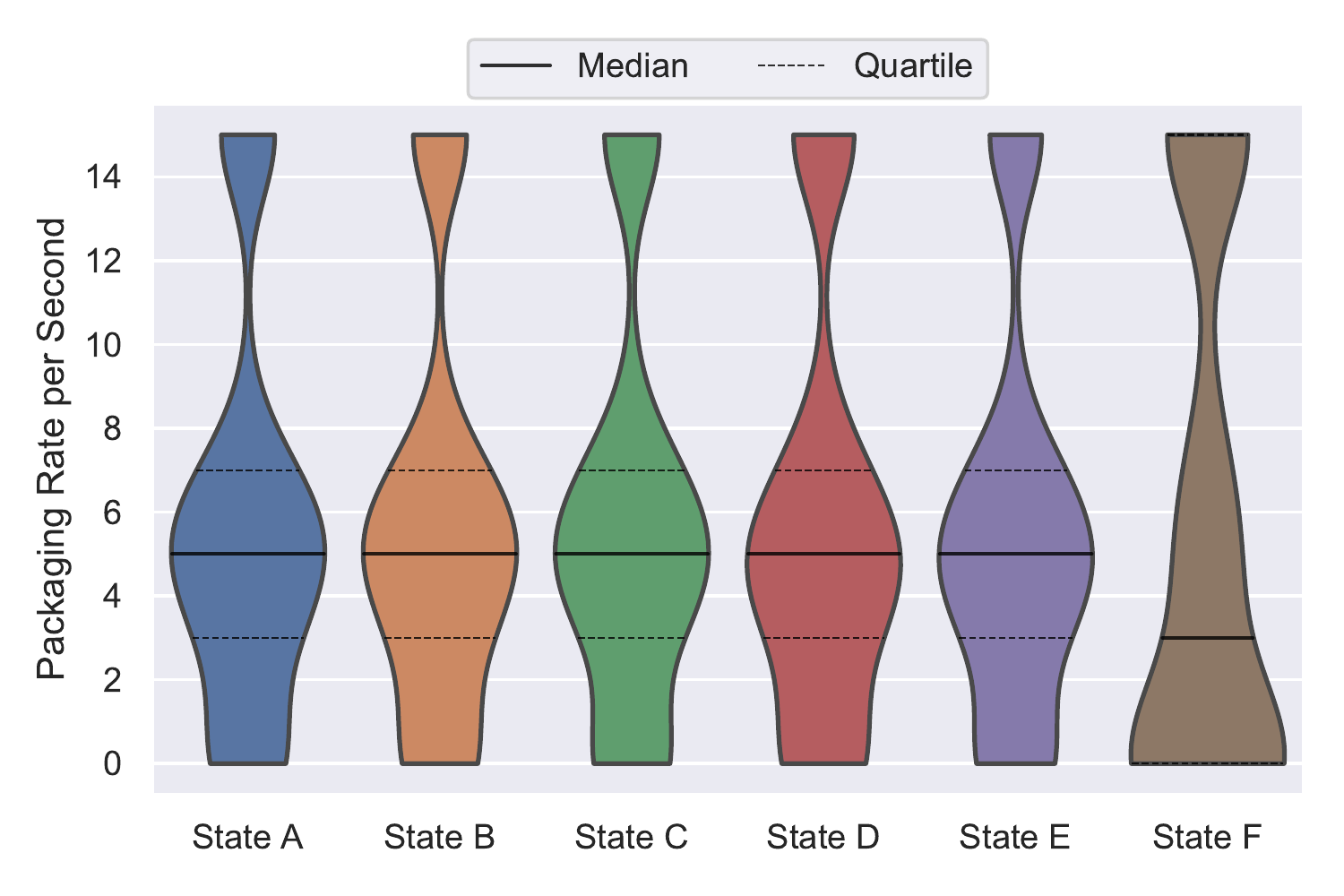}
    \caption{Distribution of packaging rate per state: When the temperature increases in state \textit{F}, \textit{packaging control} needs to pause more often resulting in more frequent packaging rates of 0 (machine is paused, 1st Quartile) and 15 (machine is running at full speed to catch up on the backlog, 3rd Quartile).}
    \label{fig:35_packaging-rate}
\end{figure}
The \textit{packaging control} reports its current packaging rate once a second.
\Cref{fig:35_packaging-rate} shows the distribution of reported values, i.e., how often each packaging rate was reported per state.
In states \textit{A}, \textit{B}, \textit{C}, \textit{D}, and \textit{E}, the workload generated by \textit{camera} and \textit{temperature sensor} is constant, so the rates are similar.
In state \textit{F}, however, the temperature sensor distributes measurements that are 30\% higher on average.
As a result, the packaging machine must halt production more frequently, i.e., the packaging rate equals zero.
This also increases the backlog; hence, the packaging machine will more frequently run at full speed to catch up on the backlog, i.e., the packaging rate equals 15.

\subsubsection{Summary} \label{sec:35_summary}

In conclusion, our experiments show that MockFog 2.0 can be used to automatically set up an emulated fog infrastructure, install application components, and orchestrate reproducible experiments.
As desired, changes to infrastructure and workload generation are clearly visible in the analysis results.
The main benefit of the MockFog approach is that this autonomous process can be integrated into a typical application engineering process.
This allows developers to automatically evaluate how a fog application copes with a variety of infrastructure changes, failures, and workload variations after each commit without access to a physical fog infrastructure, with little manual effort, and in a repeatable way~\cite{bermbach_cloud_2017}.

\section{Related Work} \label{sec:related}

Testing and benchmarking distributed applications in fog computing environments can be very expensive as the provisioning and management of needed hardware is costly.
Thus, in recent years, many approaches have been proposed that aim to enable experiments of distributed applications or services without the need for access to fog devices, especially once located at the edge.

There are several approaches that, similarly to MockFog, aim to provide an easy-to-use solution for experiment orchestration on emulated testbeds.
WoTbench~\cite{hashemian_wotbench:_2019,hashemian_contention_2020} can emulate a large number of Web of Things devices on a single multicore server.
As such, it is designed for experiments involving many power-constrained devices and cannot run experiments with many resource-intensive application components such as distributed application backends.
D-Cloud~\cite{banzai_d-cloud:_2010,hanawa_large-scale_2010} is a software testing framework that uses virtual machines in the cloud for failure testing of distributed systems.
However, D-Cloud is not suited for evaluating fog applications as users cannot control network properties such as the latency between two machines.
Héctor~\cite{behnke_hector:_2019} is a framework for automated testing of IoT applications on a testbed that comprises physical machines and a single virtual machine host. Having only a single host for virtual machines significantly limits scalability.
Furthermore, the authors only mention the possibility of experiment orchestration based on an \enquote{experiment definition} but do not provide more details.
Balasubramanian et al.~\cite{balasubramanian_rapid_2014} and Eisele et al.~\cite{eisele_towards_2017} also present testing approaches that build upon physical hardware for each node rather than more flexible virtual machines.
EMU-IoT~\cite{ramprasad_emu-iot_2019} is a platform for the creation of large scale IoT networks.
The platform can also orchestrate customizable experiments and has been used to monitor IoT traffic for the prediction of machine resource utilization~\cite{ramprasad_smart_2018}.
EMU-IoT focuses on modeling and analyzing IoT networks; it cannot manipulate application components or the underlying runtime infrastructure.

Gupta et al. presented iFogSim~\cite{gupta_ifogsim:_2017}, a toolkit to evaluate placement strategies for independent application services on machines distributed across the fog.
In contrast to our solution, iFogSim uses simulation to predict system behavior and, thus, to identify good placement decisions.
While this is useful in early development stages, simulation-based approaches cannot be used to test real application components, which we support with MockFog.
\cite{brambilla_simulation_2014, salama_iotnetsim:_2019, lera_yafs:_2019} also describe systems which can simulate complex IoT scenarios with thousands of IoT devices.
Additionally, network delays and failure rates can be defined to model a realistic, geo-distributed system.
More simulation approaches include Fog\-Ex\-plo\-rer~\cite{hasenburg_supporting_2018,hasenburg_fogexplorer_2018}, which aims to find good fog application designs, or Cisco's PacketTracer\footnote{\url{https://www.netacad.com/courses/packet-tracer}}, which simulates complex networks.
However, all these simulation approaches do not support experiments with unmodified application components.
This also applies to SimGrid~\cite{casanova_simgrid:_nodate}, a widely used framework for the simulation of distributed computer systems.
What makes SimGrid unique is its capability of simulating the communication between real application components that adhere to the MPI protocol by re-compiling them with a special toolkit.
During an experiment, each component is also emulated its own unix process.
MockFog provides stronger resource isolation with VMs and supports all TCP/IP-based application protocols.

\cite{coutinho_fogbed:_2018,mayer_emufog:_2017,peuster_medicine:_2016,peuster_containernet_2018,fontes_mininet-wifi:_2015,andres_ramiro_openleon:_2018} build on the network emulators MiniNet~\cite{de_oliveira_using_2014} and MaxiNet~\cite{wette_maxinet:_2014}.
While they target a similar use case as MockFog, their focus is not on application testing and benchmarking but on network design (e.g., network function virtualization).
Based on the papers, the prototypes also appear to be designed for single machine deployment -- which limits scalability -- while MockFog is specifically designed for distributed deployment.
Also, neither of these approaches appears to support experiment orchestration or the injection of failures.
Missing support for experiment orchestration is also a key difference between MockFog and MAMMOTH~\cite{looga_mammoth:_2012}, a large scale IoT emulator, Distem~\cite{sarzyniec_design_2013}, a tool for building experiment testbeds with Linux containers, and EmuEdge~\cite{zeng_emuedge:_2019}, and edge computing emulator that supports network replay.
Specifically in regards to our experiments in \cref{sec:exp}, this means that the application roll-out and experiment orchestration would need to be done manually. This includes updating the infrastructure, e.g., via the EmuEdge API, distributing application instructions and state changes, e.g., by writing scripts that use the curl command line utility, and monitoring transitioning conditions, e.g., by setting up a web server that collects events.
Still, one could integrate such testbed emulation approaches in MockFog's infrastructure module to support other execution environments while still benefiting from automated experiment orchestration.

OMF~\cite{rakotoarivelo_omf:_2010}, MAGI~\cite{hussain_toward_2020}, and NEPI~\cite{quereilhac_nepi:_2011} can orchestrate experiments using existing physical testbeds.
On a high level, these solutions only aim to provide functionality similar to the third MockFog module, i.e., the experiment orchestration module.

For failure testing, Netflix has released Chaos Monkey~\cite{tseitlin_antifragile_2013} as open source\footnote{\url{https://github.com/Netflix/chaosmonkey}}.
Chaos Monkey randomly terminates virtual machines and containers running in the cloud.
This approach's intuition is that failures will occur much more frequently, so engineers are encouraged to aim for resilience.
Chaos Monkey does not provide the runtime infrastructure as we do, but it would complement our approach very well.
For instance, Chaos Monkey could be integrated into MockFog's experiment orchestration module.
Another solution that complements MockFog is DeFog~\cite{mcchesney_defog:_2019}.
DeFog comprises six Dockerized benchmarks that can run on edge or cloud resources.
From the MockFog point of view, these benchmark containers are workload generating application components, i.e., load generators.
Thus, they could be managed and deployed by MockFog's application management and experiment orchestration module.
Gandalf~\cite{li_gandalf:_2020} is solely a monitoring solution for cloud deployments.
Azure, Microsoft's cloud service offer, uses Gandalf in production.
It is therefore not part of the application engineering process (\cref{fig:25_application_engineering_process}) and could be used after running experiments with MockFog.
Finally, MockFog can be used to evaluate and experiment with fog computing frameworks such as FogFrame~\cite{skarlat_framework_2018} or URMILA~\cite{shekhar_urmila:_2020}.

\section{Discussion} \label{sec:discussion}

While MockFog allows application developers to overcome the challenge that a fog computing testing infrastructure either does not exist yet or is already used in production, it has some limitations.
For example, it does not work when a specific local hardware is required, e.g., when the use of a particular crypto chip is deeply embedded in the application source code.
MockFog also tends to work better for the emulation of larger edge machines such as a Raspberry Pi but faces limitations when smaller devices are involved as they cannot be emulated accurately.

If the communication of a fog application is not based on TCP/IP, e.g., because sensors communicate via a LoRaWAN~\cite{silva_lorawan_2017} such as TheThingsNetwork\footnote{\url{https://www.thethingsnetwork.org}}, MockFog's approach of emulating connections between devices does not work out of the box as these sensors expect to have access to a LoRa sender.
With additional effort, however, application developers could adapt their sensor software to use TCP/IP when no Lora sender is available.
Furthermore, in its current state, MockFog's network manipulations also only target the application layer.
Thus, matters such as medium access contention, protocol specifications, e.g., enabling and disabling RTS/CTS for WIFI based networks, or specifics of mobile networks (4G/5G) are not emulated.
To support such use cases, others have already come up with promising solutions in the context of MiniNet that might serve as a blueprint for extending MockFog:
Fontes et al.~\cite{fontes_mininet-wifi:_2015} extended MiniNet to also emulate WIFI networks and Fiandrino et al.~\cite{andres_ramiro_openleon:_2018} added a mobile network suite to MiniNet.
Also, even wired TCP/IP connections can be affected by other users, electrical interference, or natural disasters.
For this, another solution could be to add a machine learning component to MockFog that updates connection properties based on past data collected on a reference physical infrastructure.
Still, it is hard to justify this effort for most use cases.

For the management of application containers, we decided to directly operate with Docker containers instead of using a more powerful solutions such as Kubernetes\footnote{\url{https://kubernetes.io/}}.
The main reason for this is that we do not want to assume that an application is using a certain container management solution;
especially, because in practice different solutions are used in the cloud or at the edge.
Since our MockFog prototype relies heavily on Ansible playbooks, one could easily add support for such solutions if necessary.

When an application relies on a managed Kubernetes service such as Amazon EKS\footnote{\url{https://aws.amazon.com/eks/}}, one should only use MockFog for emulating other parts of the infrastructure and set up a proxy machine that forwards traffic to EKS.
This way, the cloud part of the application can run in its regular environment and MockFog only manages the non-cloud parts by changing network characteristics between proxy and emulated edge machines.

When testing a large fog application, only the node manager is affected by increasing the number of VMs as it has to distribute instructions to more machines.
In practice, however, the node manager will even for large-scale deployments be lightly loaded as distributing instructions is not particularly resource-intensive.
Even in a situation where the node manager experiences a high load, it would only mean that state changes take slightly longer -- the application itself would not be affected at all.
It is also be possible to distribute the node manager.
In practice, though, MockFog's scalability will usually be limited by the number of VMs that can be provisioned from the cloud provider of choice.
Furthermore, it is -- simply for cost reasons -- not desirable to roll out a large-scale fog application to hundreds of MockFog nodes; it is also not necessary for testing and benchmarking purposes:
When we visualize a fog environment as a tree with the cloud as the root node and edge devices as leaves, most paths from cloud to edge will run the same application components, e.g., in the smart factory use case, there might be multiple factories that send data to the cloud.
For testing and benchmarking, however, it will usually suffice to only deploy a single example path on MockFog.
Finally, another option is to run groups of devices with similar network characteristics, such as multiple IoT sensors, on a few large VM.

\section{Conclusion} \label{sec:conclusion}

In this paper, we proposed MockFog, a system for the emulation of fog computing infrastructure in arbitrary cloud environments.
MockFog aims to simplify experimenting with fog applications by providing developers with the means to design emulated fog infrastructure, configure performance characteristics, manage application components, and orchestrate their experiments.
We evaluated our approach through a proof-of-concept implementation and experiments with a fog-based smart factory application.
We demonstrated how MockFog's features can be used to study the impact of infrastructure changes and workload variations.

\section*{Acknowledgment}

We would like to thank Elias Grünewald and Sascha Huk who have contributed to the proof-of-concept prototype of the preliminary MockFog paper~\cite{hasenburg_mockfog:_2019}.


\defbibheading{bibheading}{\section*{Bibliography}}
\printbibliography[heading=bibheading]

\end{document}